\documentclass[aps,prd,onecolumn,eqsecnum,amsmath,nofootinbib,preprintnumbers]{revtex4}%
\usepackage{color,graphicx}
\usepackage{amsfonts,amssymb,theorem,mathrsfs,times}
\usepackage{bm}
\definecolor{wred}{rgb}{0,0.618,0.0}
\definecolor{wblue}{rgb}{.0,0.0,0.618}
\definecolor{wgreen}{rgb}{.0,0.618,0.0}

\usepackage[
colorlinks=true,
filecolor=black,
linkcolor=wblue,
citecolor=blue,
pdfstartview=FitH,
pdfpagemode=None,
bookmarksopen=true,
urlcolor=blue
]{hyperref}

{\theorembodyfont{\upshape}
}
{\theorembodyfont{\upshape}
}
{\theorembodyfont{\upshape}
}
{\theorembodyfont{\upshape}
}
{\theorembodyfont{\upshape}
}
{\theorembodyfont{\upshape}
}

\newcommand{\dalm}{\kern1pt\vbox{\hrule height 0.9pt\hbox{\vrule width
0.9pt\hskip 2.5pt\vbox{\vskip 5.5pt}\hskip 3pt\vrule width
0.3pt}\hrule height 0.3pt}\kern1pt}

\begin{document}
\preprint{\hfill {\small {USTC-ICTS-16-17}}}
\title{Tunnelling phenomenon near an apparent horizon in two-dimensional dilaton gravity}

%

\author{ Li-Ming Cao$^{a,b}$\footnote{e-mail address: caolm@ustc.edu.cn},~Xiao-Zhu Jiang$^a$\footnote{e-mail address: jiangxzh@mail.ustc.edu.cn},
        ~Yuxuan Peng$^a$\footnote{e-mail address: yxpeng@mail.ustc.edu.cn}}


\affiliation{$^a$
	Interdisciplinary Center for Theoretical Study,
	University of Science and Technology of China, Hefei, Anhui 230026, China}

\affiliation{$^b$ CAS Key Laboratory of Theoretical Physics,
	Institute of Theoretical Physics,
	Chinese Academy of Sciences, Beijing 100190, China}


\begin{abstract}
Based on the definition of the apparent horizon in a general two-dimensional dilaton gravity theory, we  analyze the tunnelling phenomenon of the apparent horizon by using Hamilton-Jacobi method. In this theory the definition of the horizon is very different from those in higher-dimensional gravity theories.
The spectrum of the radiation is obtained and the temperature of the radiation is read out from this spectrum and it satisfies the usual relationship with the surface gravity.
Besides, the calculation with Parikh's null geodesic method for a simple example conforms to our result in general stationary cases.
\end{abstract}


\maketitle



\section{Introduction}
From analyzing the quantum field theory on a fixed curved background spacetime, Hawking has shown that a black hole behaves like a black body, radiating with a temperature proportional to the surface gravity of the black hole and an entropy proportional to the area of the cross section of the event
horizon~\cite{Hawking:1974rv,Hawking:1974sw}. Some nice reviews and surveys on this can be found in \cite{Parentani:1992me,Brout:1995rd,Traschen:1999zr,Jacobson:2003vx,Page:2004xp}. In Hawking's work, he suggested a heuristic picture to
explain the process of the radiation, i.e., a pair of virtual particles is created inside the horizon and the one with positive energy tunnels though the event horizon and materializes outside the horizon while the other one with negative energy is absorbed by the black hole making the mass of the black hole decrease. Then the escaped particles run to the infinity, visible to distant observers appearing as Hawking radiation~\cite{Parikh:2004ih}. Because of the radiation, the black hole loses energy and therefore shrinks, evaporating away to an unknown fate. But the actual derivations of Hawking radiation did not correspond directly to the heuristic picture~\cite{Hawking:1974rv,Hawking:1974sw,Gibbons:1976ue,Parikh:2004ih}.  And in the original derivations, a stationary black hole with an event horizon is essential.

The idea is really visual and intuitively appealing. Since the discovery of the thermal radiation, it has been widely seen as the key area to connect general classical gravity, quantum physics and thermodynamics. But unfortunately, there are two missing points in Hawking's actual derivation of the radiation. The first is the stationary black hole. With particle emission, the black hole cannot be stationary.
And further,  stationary black holes are very rare in the universe. Actual black holes are always dynamical for the accretion of matter or energy and the back-reaction of the radiation if it really exists. The second is the requirement of an event horizon which depends on the global structure of the spacetime and there are some practical issues which can not be solved easily~\cite{Ashtekar:2004cn}. So the event horizon may not exist or be a meaningless concept~\cite{Hajicek:1986hn,Ashtekar:2005cj}. Recently, a new semiclassical method to treat Hawking radiation as a tunnelling effect near the horizon has been proposed by~\cite{Kraus:1994by,Kraus:1994fj,KeskiVakkuri:1996gn,Srinivasan:1998ty,Shankaranarayanan:2000qv,Ding:2009ex}. Two principal implementations of the tunnelling approach are the null geodesic method~\cite{Parikh:1999mf} and the Hamilton-Jacobi method~\cite{Angheben:2005rm}. They apply to a large class of dynamical spacetimes with only local horizons~\cite{Padmanabhan:2003gd, Hayward:2008jq, Cai:2008gw,Banerjee:2008cf,Visser:2001kq,DiCriscienzo:2007pcr,Kerner:2007jk}. The key to the tunnelling method is the computation of the imaginary part of the single-particle action by integrating along a null path crossing the horizon.
Applying the WKB approximation, one finds that the tunnelling probability is proportional to $\exp(-2{\rm Im}S) $, where S is the classical action to the leading order in $\hbar$ (here set equal to unity).
Expanding the imaginary part of the action in terms of the particle energy, at the linear order, we get the Hawking radiation spectrum and with the quadratic term (the quadratic term appears in the computation of the null geodesic method while in that of the Hamilton-Jacobi method it seems absent), the back-reaction effect can be recovered~\cite{Parikh:1999mf,Vanzo:2011wq}.

Both the null geodesic and Hamilton-Jacobi tunnelling methods can be applied to a wide class of spacetimes and correspond very directly to the visual picture suggested by Hawking.
They are shown to be equivalent at least in the stationary case~\cite{Yang:2007zzm,Vanzo:2011wq,Wu:2007ty}. The Hamilton-Jacobi method is more powerful since it is quite convenient for treating truly dynamical black holes~\cite{Clifton:2008sb,Vanzo:2011wq}. But there is still an important missing point when one deals with the non-spherically symmetric dynamical case, i.e., the absence of an extension of Kodama-Hayward theory~\cite{Vanzo:2011wq,DiCriscienzo:2007pcr}.
In axis-symmetric dynamical spacetimes, a method called generalised tortoise coordinate transformation (GTCT) is developed by authors in~\cite{Jing1996,Jing:1997ui} and the Hawking radiation is related to the event horizon.
In spherically symmetric cases, the Kodama vector~\cite{Kodama:1979vn} instead of the original Killing vector can be used to define the direction of the Hamilton flow and the horizons would be some quasi-local horizons whose definitions closely rely on the concept of Hayward's trapping horizon~\cite{Hayward:1997jp,Hayward:1998jj}.
Recently, Senovilla and Torres provided a general formula which could be used to analyze the phenomenon of tunnelling in arbitrary spacetimes with marginally trapped surfaces (MTSs) by using the dual expansion vector to replace the Kodama vector~\cite{Senovilla:2014ika}. And in this paper, we will use this method to analyze the tunnelling phenomenon in a general two-dimensional dilaton gravity.

Some important technical complications in dealing with the basic questions of quantum gravity in higher dimensions make the treatment of these basic questions extremely difficult.
Therefore, a rich literature has been developed on lower-dimensional models of gravity in the past years.
For instance, the two-dimensional dilaton gravity has been widely studied over the past twenty years. This kind of gravity can be obtained from the spherically symmetric reduction of Einstein gravity theory in higher dimensions or by eliminating the Weyl anomaly on string world sheet, e.g. the Callan, Giddings, Harvey and Strominger (CGHS) model~\cite{Callan:1992rs} and the generalised dilaton theories (GDTs) in two dimensions(for a nice review, see~\cite{Grumiller:2002nm}).
A lot of black hole solutions and cosmological solutions have been found in these theories and most of them become general dynamical solutions when some matter fields are introduced which will make the situations become complicated.
In the paper~\cite{cao-cai}, the definition of an apparent horizon is provided in the general two-dimensional  dilaton gravity theory and then the mechanics of the horizon is constructed by introducing a quasi-local energy and a Kodama-like vector field.
As we can see, their definition of the horizon is very general and quasi-local, therefore, it is very interesting and important to make sure whether the radiation still exists or not and to study the properties of the radiation if it exists.

This paper is organized as follows: In Sec.II, we briefly review the definition of the apparent horizon in the general two-dimensional dilaton gravity and give some simple discussions of its properties.
In Sec.III, we use the Hamilton-Jacobi method to analyze the tunnelling phenomenon in the general cases in the two-dimensional dilaton gravity. We then use the null geodesic method to deal with a specific case and compare the results of the two situations in Sec.IV.
Finally, the conclusions and discussions are in Sec.V. We will use the following conventions in this paper: the constants $c$,~$G$,~$k_B$ and $\hbar$ set equal to unity. Latin indices as $a\,,b$ are used to denote the abstract indices and Greek indices as $\mu\,,\nu$ are used to denote the components of a tensor.

\section{apparent horizon in two-dimensional dilaton gravity }
In this section we will follow the discussions in~\cite{cao-cai}. The action of a general two-dimensional dilaton gravity can be expressed as ~\cite{Grumiller:2002nm}
\begin{equation}\label{action}
  I =\int {\rm d^2 }x\sqrt{-h}\left[ \Phi R +U(\Phi)D^a\Phi D_a\Phi +V(\Phi) +\mathscr{L}_m  \right] \, ,
\end{equation}
where $h_{ab}$ is the spacetime metric, $h$ its determinant, $\Phi$ the so-called dilaton field, $R$ the Ricci scalar, and $\mathscr{L}_m$ is the matter Lagrangian. The equation of motion for the dilaton field $\Phi$ is
\begin{equation}
 \label{dilatoneom}
  R -U'(\Phi)D_a\Phi D^a\Phi +V'(\Phi) -2U(\Phi)\Box\Phi +\mathscr{T}_m =0 \, ,
\end{equation}
where $D_a$ is the derivative operator compatible with $h_{ab}$, and $\Box =D_aD^a$. The prime stands for the derivative with respect
to $\Phi$: ${\rm d/d}\Phi$, while the scalar $\mathscr{T}_m$ is defined as
\begin{equation}
  \mathscr{T}_m :=\frac{\partial\mathscr{L}_m}{\partial\Phi} -D_a\frac{\partial\mathscr{L}_m}{\partial(D_a\Phi)} +\cdots  \, .
\end{equation}
The equation of motion for the metric $h_{ab}$ is
\begin{equation}
 \label{eommetric}
 U(\Phi)D_a\Phi D_b\Phi-\frac{1}{2}U(\Phi)D^c\Phi D_c\Phi h_{ab} -D_{a}D_{b}\Phi
  +\Box \Phi h_{ab}-\frac{1}{2}V(\Phi)h_{ab}= T_{ab}\, ,
\end{equation}
where $T_{ab}$ is the energy momentum tensor of the matter field.

Assume $\{\ell^a,n^a \}$ is a null frame in the spacetime and the metric can be expressed as
\begin{equation}
\label{null frame}
  h_{ab} =-\ell_a n_b -n_a \ell_b \, ,
\end{equation}
while $\ell^a$ and $n^a$ are two future directed null vector fields which are globally defined on the spacetime and satisfy $\ell_a n^a =-1$. Also we assume $\ell_a$ and $n^a$ are outer pointing and inner pointing respectively. Now we introduce an important vector field $\phi^a =D^a\Phi$. In the null frame, it is easy to see
\begin{equation}
 \phi_a\phi^a =D_a\Phi D^a\Phi =-2\mathcal{L}_\ell\Phi \mathcal{L}_n\Phi \, .
\end{equation}
Considering the causality of this vector field, the spacetime can be divided into several parts, and in each part the vector field $\phi^a$ is either spacelike or timelike and on the boundary of any part it is null, and this boundary can be defined as a kind of horizon.
The definitions are~\cite{cao-cai}
\begin{description}
\label{definition of horizon}\label{define horizon}
  \item[future outer horizon :] $ \mathcal{L}_\ell\Phi =0\, , \mathcal{L}_n\Phi<0 \,  ,\mathcal{L}_n\mathcal{L}_\ell\Phi<0 $ \, ;
  \item[future inner horizon :] $ \mathcal{L}_\ell\Phi =0\, , \mathcal{L}_n\Phi<0 \, , \mathcal{L}_n\mathcal{L}_\ell\Phi>0 $ \, ;
  \item[past outer horizon :]~~~$ \mathcal{L}_n\Phi =0\, , \mathcal{L}_\ell\Phi>0 \, , \mathcal{L}_\ell\mathcal{L}_n\Phi>0 $ \, ;
  \item[past inner horizon :]~~~$ \mathcal{L}_n\Phi =0\, , \mathcal{L}_\ell\Phi>0 \, , \mathcal{L}_\ell\mathcal{L}_n\Phi<0 $ \, .
\end{description}
From the definition we can see that the horizon is in fact a trapping horizon, but we will use the term ``apparent horizon" as usual in the two-dimensional dilaton gravity.
Below we will focus on future outer horizons.

From the discussion in the introduction we can see that the radiation or particle creation is related to some local region around which some key vector field changes its causality from temporal to spatial. The key vector field is the Killing vector field in stationary spacetimes or the Kodama vector field in spherically symmetric dynamical spacetimes while in more general spacetimes which are not spherically symmetric is the dual expansion vector field:
\begin{equation}
 H^a \equiv -\theta_n\ell^a +\theta_\ell n^a \, ,
\end{equation}
where $\theta_\ell$ and $\theta_n$ are the expansion scalars of the outgoing and ingoing null congruences normal to a codimension-2 spacelike surface respectively~\cite{Senovilla:2014ika}.
At first sight, this vector field cannot be defined in the general two-dimensional dilaton gravity theory since the codimension-2 surface shrinks to a point in two dimensions.
However, thanks to the dilaton field $\Phi$, if we regard $\mathcal{L}_\ell\Phi$ and $\mathcal{L}_n\Phi$ as the counterparts in two dimensions of $\theta_\ell$ and $\theta_n$ respectively, we can define the dual expansion vector field in this case as
\begin{equation}\label{dual expansion vector}
 H^a \equiv -(\mathcal{L}_n\Phi)\ell^a +(\mathcal{L}_\ell\Phi) n^a \, ,
\end{equation}
which is dual to the vector field $\phi^a$.
The dual expansion vector field $H^a$ is also parallel to the Kodama vector field
$K^a$ (\ref{Kodama Relation}) which is used to define the surface gravity in~\cite{cao-cai}.
The region with $\mathcal{L}_\ell\Phi<0$ and $\mathcal{L}_n\Phi<0$ can be called the trapped region of the spacetime, and the boundary of the region therefore can be defined as a horizon.
From this point of view, the definition of the horizon is natural and reasonable. Also, the future outer horizon has the traditional meaning, i.e. the boundary of a region from which the light cannot escape along a classical trajectory. With the expression of the dual expansion vector
(\ref{dual expansion vector}), we can see that
\begin{equation}
\label{dull causality}
   h_{ab} H^a H^b =2\theta_\ell \theta_n =2\mathcal{L}_\ell\Phi \mathcal{L}_n\Phi \, .
\end{equation}
So the dual expansion vector is timelike outside the trapped region, spacelike inside the trapped region and null on the horizon, which is the desirable property of the dual expansion vector field.

\section{general tunnelling process}
In this section we will use the Hamilton-Jacobi tunnelling method to analyze the tunnelling phenomenon associated to the future outer horizon in the two-dimensional dilaton gravity theory. First of all, we should make sure that, the WKB approximation is still justified near the horizon. We choose a family of fiducial observers $\xi^a$. When they approach the horizon, their 4-velocity $\xi^a$ becomes
\begin{equation}
  \xi^a = \frac{H^a}{\sqrt{-H^b H_b}} \, .
\end{equation}
In Schwarzschild spacetime, they are the static observers outside the horizon. They measure an energy for the particle $\tilde{\omega}=-\xi^a D_a S$, where S is the classical action of the particle, to the leading order of $\hbar$. This energy becomes infinity near the horizon as the dual expansion vector $H^a$ is null on the horizon, making the WKB approximation fully reliable, at least in this reference frame.

The WKB approximation tells us that the tunnelling probability of particles along a classically forbidden trajectory from inside to the outside of the horizon which is regarded as a potential barrier is
\begin{equation}
\label{WKB}
  \Gamma \propto \exp(-{\rm 2 Im} S) \, ,
\end{equation}
where S is the classical action of the massless particle, to leading order in $\hbar$. And the action can be calculated by
\begin{equation}
\label{integration}
  S =\int {\rm d}S \, .
\end{equation}
As we need only the imaginary part of the action, the integration path should be chosen to be a special one that can cross the horizon. Since then, the integration will give an imaginary contribution to the action. We also assume that the particle's action satisfies the relativistic Hamilton-Jacobi equation,
\begin{equation}
\label{Hamilton-Jacobi equation}
  h^{ab} D_a S D_b S =0 \, .
\end{equation}

In the general spacetime of the two-dimensional dilaton gravity theory, we can define a Kodama-like vector field as
\begin{equation}
\label{Kodama vector}
  K^a =-e^Q\epsilon^{ab}D_b \Phi \, ,
\end{equation}
where $Q$ is defined by $Q'(\Phi) =-U(\Phi)$ and $\epsilon_{ab}$ is the volume element of the two-dimensional spacetime. After some simple calculation, given the expression (\ref{dual expansion vector}), we have
\begin{equation}
\label{Kodama Relation}
  K^a =e^Q H^a \, ,
\end{equation}
which is similar to the so-called generalised Kodama vector field $\sqrt{A(\mathcal{S})/16\pi}H^a$ (where $\mathcal{S}$ is a codimension-2 spacelike surface and $A(\mathcal{S})$ is its area) in~\cite{Senovilla:2014ika} except for the coefficient of the dual expansion vector as there is no codimension-2 surface in this case. At first sight, the coefficient should be vanishing as the codimension-2 surface becomes a dot now. However, the fact is no. With this Kodama-like vector field, one can define an energy of the tunnelling particles
\begin{equation}
\label{particle energy}
  \omega =-K^a D_a S \, .
\end{equation}

Given the double null metric (\ref{null frame}) and taking (\ref{Hamilton-Jacobi equation}), (\ref{Kodama Relation}), (\ref{particle energy}) into consideration together, we get
\begin{eqnarray}
 0 &=&(\ell^a\partial_a S)(n^b\partial_b S)\, ,    \\
 \omega &=&e^Q(\mathcal{L}_n\Phi)(\ell^a D_a S) -e^Q(\mathcal{L}_\ell\Phi)(n^a D_a S)  \,,
\end{eqnarray}
from which we can get two solutions $\ell^a D_a S =0$ or $n^a D_a S =0$. We need only the solution which corresponds to the case where the particles can cross the horizon i.e. the solution of the outgoing mode
\begin{eqnarray}
	\label{outgoing solution1} \ell^a D_a S &=& 0  \, , \\
	\label{outgoing solution2} n^a D_a S &=& -\frac{\omega}{e^Q(\mathcal{L}_\ell\Phi)} \, .
\end{eqnarray}
At first sight, we may regard this as the ingoing mode for the derivative of the action with respect to the ingoing null curve, but in fact, the integral will be carried along the inverse direction of the ingoing null curve. Also, due to the characteristics of the horizon~\cite{cao-cai}, the outgoing null curve cannot cross the horizon i.e., its coordinate is not regular across the horizon so that the integration cannot be done along that path.
Whereas, moving along the inverse direction of the ingoing curve means that the particles must travel back in time, which is classically forbidden~\cite{Parikh:1999mf}, and this classically forbidden integration will contribute to the imaginary part of the action. For our choice, we get the fact that $\mathcal{L}_\ell\Phi \rightarrow 0$ when approaching the horizon which informs us that $n^a D_a S$ diverges as the particles cross
the horizon and this is just what we want.

Insert (\ref{outgoing solution1}) and (\ref{outgoing solution2}) into (\ref{integration}), the imaginary part of the action for an outgoing particle tunnelling from a point ({\em in}) inside the horizon to a point ({\em out}) outside the horizon can be expressed as
\begin{eqnarray}
  {\rm Im} S &=& {\rm Im}\int_{in}^{out} {\rm d} S \\
  {} &=& {\rm Im}\int_{in}^{out} -\left[ (n^\nu\partial_\nu S)\ell_\mu{\rm d}x^\mu +(\ell^\nu\partial_\nu S)n_\mu{\rm d}x^\mu \right] \\
 \label{imaginary part} {} &=& {\rm Im}\int_{in}^{out}\frac{e^{-Q}\omega}{\mathcal{L}_\ell\Phi} \ell_\mu{\rm d}x^\mu \, .
\end{eqnarray}
As we can see, the integration is divergent. We will regularise the divergence according to Feynman's $i\epsilon-$ prescription. Denote by $\lambda$ the parameter of the tangent vector of the ingoing null curve i.e. $n^a=({\rm d}/{\rm d}\lambda)^a$, then the integration 1-form can be expressed as
\begin{equation}
\label{integration 1-form}
  \ell_\mu {\rm d}x^\mu = -{\rm d}\lambda  \, .
\end{equation}
From the discussion above, only a small segment crossing the horizon will make sense. And recall that $\mathcal{L}_\ell\Phi=0$ on the horizon, so in the neighbourhood of the horizon it can be expressed as
\begin{equation}
\label{near horizon}
 \mathcal{L}_\ell\Phi \approx \left. \frac{{\rm d}(\mathcal{L}_\ell\Phi)}{{\rm d}\lambda} \right|_H(\lambda -\lambda_0) \, ,
\end{equation}
where $\lambda_0$ is the intersection point of the curve with the horizon. Now, inserting (\ref{integration 1-form}) and (\ref{near horizon}) into
 (\ref{imaginary part}) and using the Feynman's $i\epsilon-$ prescription, the imaginary part of the action is (for more details one can refer to \cite{Vanzo:2011wq})
\begin{equation}
  {\rm Im}S =-{\rm Im}\int_{in}^{out} \frac{\omega e^{-Q}}{\left. {\rm d}(\mathcal{L}_\ell\Phi)/{\rm d}\lambda \right|_H (\lambda -\lambda_0 -i\epsilon)}
   {\rm d}\lambda
            =\frac{\pi \omega}{\kappa}   \, ,
\end{equation}
where $\kappa$ is defined as
\begin{equation}
  \kappa =\left.-e^Q\frac{{\rm d}(\mathcal{L}_l\Phi)}{{\rm d}\lambda}\right|_H =-e^Q\mathcal{L}_n \mathcal{L}_\ell\Phi |_H \, .
\end{equation}
From the discussion in~\cite{cao-cai}, we have the equation
\begin{equation}
  \mathcal{L}_n \mathcal{L}_\ell\Phi =-\kappa_{(n)}(\mathcal{L}_\ell\Phi) -(1/2)\Box\Phi \, ,
\end{equation}
where $\kappa_{(n)}$ is a scalar defined by $\kappa_{(n)}=-\ell_a n^b D_b n^a$. As we treat the future outer horizon here, we have
\begin{equation}
  \Box\Phi >0 \,.
\end{equation}
Then $\kappa$ can be expressed as
\begin{equation}\label{surfacegra}
  \kappa =\frac{1}{2}e^Q\Box\Phi \, ,
\end{equation}
which is nothing but the surface gravity in~\cite{cao-cai} and always positive on the future outer horizon. Meanwhile, the tunnelling probability (\ref{WKB}) becomes
\begin{equation}
  \Gamma \propto \exp(-\frac{2\pi\omega}{\kappa}) \,.
\end{equation}
Combining with the Boltzmann factor $\exp (-\omega /T)$ of the thermal radiation, we get
\begin{equation}
  T =\frac{\kappa}{2\pi} \,,
\end{equation}
which is exactly the famous relation between temperature and surface gravity in the Hawking radiation.

Therefore, we have proven that the tunnelling phenomenon still exists in the two-dimensional dilaton gravity theory as long as there is a local horizon in the spacetime. The temperature of the radiation and the surface gravity of the horizon satisfy the usual identity.

\section{a simple example}
In this section, we will use the null geodesic method to recalculate the radiation spectrum. As the null geodesic method is very inconvenient to treat truly dynamical black holes~\cite{Vanzo:2011wq}, we consider the vacuum solution of (\ref{eommetric}). In Eddington-Finkelsten gauge, a general solution has a simple form~\cite{cao-cai}
\begin{equation}
\label{EFmetric}
  {\mathrm ds}^2 =-e^Q(w-2m) {\rm d}v^2 +2{\rm d}v {\rm d}r   \,,
\end{equation}
where functions $Q(r)$, $w(r)$ and $r$ are defined by
\begin{equation}
\label{Qwr}
 U=-Q' \,  ,V=e^{-Q}w' \, ,{\rm d}r=e^Q{\rm d}\Phi \,,
\end{equation}
and $m$ is the black hole mass. When one uses the null geodesic method, it is usually convenient to choose the  Painlev\'e-Gullstrand gauge~\cite{Painlev}. If we use a function $f$ to denote the metric component $h_{vv}$ then the Painlev\'e time $t$ can be defined by
\begin{equation}
  {\rm d}t ={\rm d}v -f^{-1}\left(1- \sqrt{1-f} \right) {\rm d}r  \,.
\end{equation}
The solution (\ref{EFmetric}) can be expressed as
\begin{equation}
  \label{PGmetric}
  {\rm ds}^2 =-f{\rm d}t^2 +2\sqrt{1-f}{\rm d}t{\rm d}r +{\rm d}r^2 \, ,
\end{equation}
which is a stationary metric and the apparent horizon is defined by
\begin{equation}
  \label{PGhorizon}
  f=e^Q(w-2m)=0 \, ,
\end{equation}
which can be regarded as a special case of the general definition in Sec.II. In this stationary spacetime, the Kodama-like vector field (\ref{Kodama vector}) becomes
\begin{equation}
  K^a =\Big(\frac{\partial}{\partial t}\Big)^a \, ,
\end{equation}
which is nothing but the Killing vector field of the spacetime. The ``energy" (\ref{particle energy}) now is invariant along the particle's propagating path, and it is just the particle energy measured by the observers at infinity when the spacetime is asymptotically flat. The outgoing null geodesic is given by
\begin{equation}
  \label{null geodesic}
  \dot{r} =1-\sqrt{1-f} \, ,
\end{equation}
where the overdot stands for the derivative with respect to $t$. The imaginary part of the action of an outgoing particle crossing the horizon can be expressed as
\begin{eqnarray}
  {\rm Im} S &=&   {\rm Im}\int_{in}^{out} p_r{\rm d}r ={\rm Im}\int_{r_{in}}^{r_{out}} \int_0^{p_r}{\rm d}p'_r{\rm d}r \,,
\end{eqnarray}
where the prime is a symbol of integration variable which should not be confused with the derivative with respect to $\Phi$.
In order to calculate the integral, we change the integration variable from momentum to energy by using the Hamilton's equation
$\dot{r}=\left.{\rm d}H/{\rm d}p_r\right|_r$, where $H$ is the generator of Painlev\'e time~\cite{Parikh:2004rh}. Then we get
\begin{eqnarray}
  {\rm Im}S &=& {\rm Im}\int_{m}^{m-\omega}\int_{r_{in}}^{r_{out}}\frac{{\rm d}r}{\dot{r}} {\rm d}H \\
  {} &=& {\rm Im}\int_0^{\omega} \int_{r_{in}}^{r_{out}} \frac{{\rm d}r}{1-\sqrt{1-e^Q(w-2m+ 2\omega')}} (-{\rm d}\omega') \\
  {} &=& {\rm Im} \int_{r_{in}}^{r_{out}} \frac{(-\omega){\rm d}r}{1-\sqrt{1-e^Q(w-2m)}} +O(\omega^2) \\
  {} &=& {\rm Im} \int_{\Phi_{out}}^{\Phi_{in}} \frac{\omega e^Q{\rm d}\Phi}{1-\sqrt{1-e^Q(w-2m)}} +O(\omega^2) \,.
\end{eqnarray}
The third step comes from the fact that $\omega\ll m$ and we use the definition ${\rm d}r =e^Q{\rm d}\Phi$ in the last step. Since we do not know the exact expression of the functions $Q$ and $w$, we cannot get the precise result of the integration. To the first order in $\omega$, the integration also has a pole and we can regularise its divergence with the Feynman's $i\epsilon-$ prescription as before. The final result is
\begin{equation}
  \label{result back-reaction}
  {\rm Im}S = \frac{2\pi\omega}{w'} +O(\omega^2) \,,
\end{equation}
where the prime stands for the derivative with respect to $\Phi$ as (\ref{Qwr}). Comparing with the Boltzmann factor again, we get an $\omega^2$ correction term to the radiation spectrum due to the inclusion of back-reaction effect~\cite{Parikh:1999mf}.
Neglecting the correction term in the radiation spectrum, we recover the radiation and the radiation temperature can be expressed as
\begin{equation}
  T =\frac{w'}{4\pi} \, .
\end{equation}
Taking the trace of (\ref{eommetric}) in the vacuum case gives $\Box\Phi =V$ and considering the equation (\ref{surfacegra}) and the definition (\ref{Qwr}), we have that
\begin{equation}
T = \frac{\kappa}{2\pi}\,,
\end{equation}
which agrees with the general result in the previous section.


As we can see, in the stationary case, the Kodama-like vector field $K^a$ becomes the Killing vector field of the spacetime and when the spacetime is asymptotically flat the relative energy $\omega$ becomes the particle energy measured by the observers at infinity. So the radiation associated to the horizon is exactly the Hawking radiation and the temperature $T$ is the temperature of the Hawking radiation.

\section{conclusions}
In this paper, we have used the Hamilton-Jacobi tunnelling method to analyze the tunnelling phenomenon in a general two-dimensional dilaton gravity theory. We have shown that, the radiation still exists in the general spacetimes and can be associated to the apparent horizon.
We also obtained the spectrum of this radiation.
The temperature of the radiation has been read out from this spectrum and it satisfies the usual relationship with the surface gravity.
These are universal conclusions for all the GDTs and a large class of CGHS models and any other two-dimensional dilaton gravity theories derived from the action like (\ref{action}). We also used the null geodesic method to analyze the stationary case of the theory and got the same result.
This suggests that, the tunnelling phenomenon is a universal phenomenon of the future outer horizon, independent of the effects of ``large isometries" \cite{Vanzo:2011wq}, which is the same as the case in higher dimensions.
With the null geodesic method, we have found the correction term to the radiation spectrum in $\omega^2$ order which was not seen in the Hamilton-Jacobi method, though the equivalence of the two methods has been proven for stationary black holes in higher dimensions as mentioned in the introduction before with references there. This issue may be further investigated in the future.

\section{acknowledgements}
This work was supported in part by the National Natural Science Foundation of China with grants No.11622543 and No.11235010. This work was also Supported by the Open Project Program of Key Laboratory of Theoretical Physics, Institute of Theoretical Physics, Chinese Academy of Sciences, China (No.Y5KF161CJ1). LMC would like to thank Rong-Gen Cai for his useful discussion and comments.


\end{document}